\documentclass[aps,prb,showpacs,twocolumn]{revtex4-1}
\usepackage{graphicx}
\begin{document}
%
%
\title{Variation of the extended $s$-wave superconducting order parameter: from s-wave to g-wave}
%

\author{Heesang Kim}
\affiliation{Department of Physics, Soongsil  University, Seoul
156-743, Korea}
\affiliation{Institute for Integrative Basic Sciences, Soongsil  University, Seoul
156-743, Korea}
\author{Hyunhee Chung}
\affiliation{Department of Physics, Soongsil  University, Seoul
156-743, Korea}
\author{Nammee Kim}
\email{nammee@ssu.ac.kr} \affiliation{Department of Physics,
Soongsil  University, Seoul 156-743, Korea}

\date{\today}
\begin{abstract}
We investigate evolution of properties of an extended $s$-wave superconductor, when the order parameter varies from an $s$-wave to a $g$-wave continuously, by using a model order parameter $\Delta(\hat{k}) = \Delta_{0}((1-x)+x \sin^{4}\theta \cos 4\phi)$. The evolution of the gap amplitude, the density of states, and the specific heat are mainly focused on. For $x<0.5$, due to the existence of a finite sized gap, the characteristic behaviors more or less follow those of the $s$-wave. Sudden changes in the characteristic behaviors come out for $x \ge 0.5$, due to appearances of nodes. For $x=0.5$, point nodes in the order parameter on the Fermi surface appear, while for $x>0.5$, line nodes appear. Although they are different kinds of nodes which would usually induce different power-law dependencies in superconducting properties, interestingly enough, they give rise to the same characteristic behavior. The detailed structure of the point nodes for $x=0.5$ is investigated, and it is explained why they lead to the same dependence as the line nodes.
\end{abstract}

\maketitle
%
Symmetry of a superconducting order parameter is always of great
interest, since it is directly related to the shape of the
interaction responsible for the cooper pair formation.
Superconductivity in strongly correlated systems, such as cuprates
and heavy fermion systems, usually appears with a highly
anisotropic pairing, whose order parameter has strong
$\hat{k}$-dependence and nodes.\cite{hardy,harlingen,tsuei,stewart,sauls,ott} Phonon-mediated superconductivity
appears with an $s$-wave or relatively less anisotropic pairing.
However, it has been reported that the existence of Fermi
surface nesting may lead to strong anisotropy of order parameter
even in phonon-mediated superconductors such as
$\mbox{Y}\mbox{Ni}_{2}\mbox{B}_{2}\mbox{C}$ and
$\mbox{Lu}\mbox{Ni}_{2}\mbox{B}_{2}\mbox{C}$.\cite{nohara,tuson,canfield} The strong
$\hat{k}$-dependence may result in nodes as well, and yet the
order parameter keeps the full rotational symmetry of the host
metal in this case unlike the $d$-wave in the cuprates. This
anisotropic order parameter transforms according to the totally
symmetric representation in the group theoretical point of view,
and can be classified as an $s$-wave. It is often called `an
extended $s$-wave'.\cite{soin,hirsch,hskim01,kuro} $s+g$-wave order parameter, studied in
connection with the non-magnetic borocarbides,\cite{iza,maki01,maki02}
is a good example of the extended $s$-wave.

In this paper, the effect of variation of the gap anisotropy on
superconducting properties of the $s+g$-wave superconductor is
studied as a concrete example of the extended $s$-wave order
parameter. Starting from an $s$-wave, and adding the $g$-wave
component, we investigate evolution of the gap anisotropy and nodes,
change of the maximum and minimum of the gap,
temperature dependence of the gap amplitude, the density of
states, and the specific heat.

The $s+g$-wave order parameter is set as the following form.
\begin{equation}
\Delta(\hat{k}) = \Delta_{0} ( (1-x) + x \sin^{4} \theta \cos 4\phi).
\label{eq_op}
\end{equation}
%
%
\begin{figure}
\includegraphics[width=8.0cm]{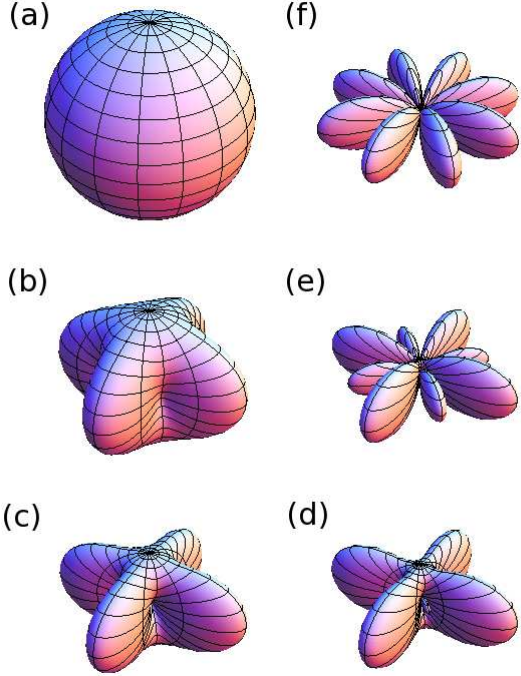}
\caption{\label{fig:op} Variation of the order parameter from an $s$-wave to a $g$-wave. $\Delta(\hat{k}) / \Delta_{0} =( (1-x) + x \sin^{4} \theta \cos 4\phi)$ for $x=0$ in (a), $x=0.3$ in (b), $x=0.5$ in (c), $x=0.65$ in (d), $x=0.85$ in (e), $x=1$ in (f).}
\end{figure}
%
Figure 1 shows the variation of the superconducting order parameter,
{\it i.e.}, the excitation energy gap of quasiparticles, from an $s$-wave to a $g$-wave, by increasing weighting
parameter $x$ from 0 to 1. Figure 1(a) simply demonstrates an $s$-wave.
The gap develops an anisotropy and the anisotropy grows  with parameter $x$ being increased, as shown in Fig. 1(b) when $x=0.3$.
There is no phase change yet, since the $s$-wave component still dominates over the $g$-wave component. In Fig. 1(c), when $x=0.5$, $s$-wave and $g$-wave components have the equal weights, and there emerge four accidental point nodes along the equator. In Figs. 1(d) and 1(e), weight of the $g$-wave part exceeds that of the $s$-wave.
As parameter $x$ increases further, there grow small paddle-like pieces of order parameter where the point nodes have been located. These pieces have phase difference by $\pi$ from the rest of the order parameter. In other words, they have different sign. Line nodes are naturally expected at the boundaries of these regions on the Fermi surface. Figure 1(f) exhibits an $g$-wave. The paddle-like regions grow equal to the rest of the order parameter, which results in eight line nodes along the azimuthal angle and a point node at each pole.

The gap amplitude of quasiparticle's excitation energy, $\Delta_{0}$, is obtained by solving the gap equation.
\begin{equation}
\frac{1}{g} = N(0)\pi T \sum_{n}^{'} \bigg< \frac{\big| f(\hat{k}) \big|^{2}}{\sqrt{\omega_{n}^{2} + \big|\Delta(\hat{k})\big|^{2}}} \bigg>_{FS}.
\end{equation}
This is the weak-coupling gap equation, where $g$ is the coupling constant, $N(0)$ is the normal state density of states at the Fermi surface, and $\omega_{n} = (2n+1)\pi T$ is the Matsubara frequency. The Matsubara sum needs a cut-off, and the bracket represents taking an average over the Fermi surface. Here, the natural unit is used so that $k_{B}=\hbar=c=1$.
Notice that $f(\hat{k})$ is a normalized basis function of a certain group representation such that
$\Delta(\hat{k}) = \Delta_{n} f(\hat{k})$ and $\Big< \big| f(\hat{k}) \big| ^{2} \Big>_{FS} = 1$.
Since the functional form in Eq. (\ref{eq_op}) is not normalized,
$f(\hat{k}) = ( (1-x) + x \sin^{4} \theta \cos 4\phi)/\sqrt{(256-512x+291x^{2})\pi/512}$ and the normalized gap amplitude would be
$\Delta_{n} = \Delta_{0} \sqrt{(256-512x+291x^{2})\pi/512}$.
The coupling constant $g$ is not an easy parameter to determine experimentally. In calculations, the parameter $g$ along with the normal state density of states $N(0)$ can be replaced by a single input parameter $T_{c}$, the transition temperature. Then, the gap equation can be written as
\begin{equation}
\psi \left( \frac{E_{c}}{2\pi T_{c}} +1 \right) - \psi\left(\frac{1}{2}\right) = 2\pi T \sum_{n \geq 0}^{'} \bigg< \frac{\big| f(\hat{k}) \big|^{2}}{\sqrt{\omega_{n}^{2} + \big|\Delta(\hat{k})\big|^{2}}} \bigg>_{FS},
\end{equation}
where $\psi(x)$ is the digamma function and $E_{c}$ is the energy cut-off. Notice that the solution of this gap equation dose not depend on the choice of a big enough cut-off energy in the weak coupling limit.\cite{gian,nicol}

%
%
\begin{figure}
\includegraphics[width=8.0cm]{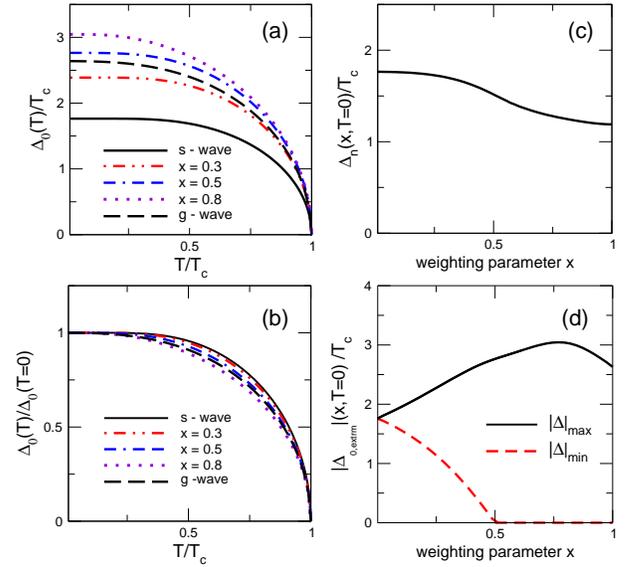}
\caption{\label{fig:gap} (a) Gap amplitude as a function of temperature for various weighting parameters $x$. (b) Gap amplitude normalized by its zero temperature value as a function of temperature. (c) Normalized gap amplitude $\Delta_{n}$ as a function of weighting parameter $x$ at zero temperature. Note that $\Delta(\hat{k})=\Delta_{n} f(\hat{k})$, where $\big< \big| f(\hat{k}) \big| ^{2} \big>_{FS} = 1$, and therefore $\Delta_{n} = \Delta_{0} \sqrt{(256-512x+291x^{2})\pi/512}$. (d) Extremum values of the gap function as a function of weighting parameter $x$ at zero temperature. Note that $|\Delta|_{\max}=\Delta_{0}$ and $|\Delta|_{\min} = \max\{ \Delta_{0}(1-2x),0\}$.}
\end{figure}
%
Figure 2(a) shows the gap amplitude $\Delta_{0}(T)$ as a function of temperature for various choices of weighting parameter $x$. The solid curve in Fig. 2(a) represents the $s$-wave case, namely, the BCS curve. As the weighting parameter $x$ increases, the gap amplitude also increases up to $x \sim 0.8$ and it starts to decrease after then. The temperature dependence of the gap amplitude seems pretty much to follow that of the BCS curve. However, a closer look tells us that the temperature dependence of the gap amplitude normalized by its zero temperature value, $\Delta_{0}(T)/\Delta(0)$, deviates from the BCS curve. It is shown in Fig. 2(b). We note that all the curves fall below the BCS curve, and it deviates the most when $x \sim 0.8$. The degree of deviation seems commensurate with the value $\Delta_{0}(T=0)$, although the value does not have to be related to the degree of deviation.
Figure 2(c) shows the normalized gap amplitude $\Delta_{n}$ at zero temperature as a function of weighting parameter $x$. Notice that $\Delta_{n}$ is different from $\Delta_{0}$ which appears in Eq. (\ref{eq_op}). Since $\Delta_{n}$ is defined by using a normalized basis function, this is the quantity related to the condensation energy of the superconducting state rather than $\Delta_{0}$. The normalized gap amplitude $\Delta_{n}$ deceases monotonously as the weighting parameter $x$ increases. This observation implies that the anisotropy in the superconducting order parameter tends to suppress the condensation energy.
Figure 2(d) shows extremum values of the gap function at zero temperature as a function of the weighting parameter $x$, $|\Delta_{\rm{0, extrm}}(x, T=0)|/T_{c}$. The solid curve corresponds to the gap maxima $|\Delta|_{\max}/T_{c}$, while the dashed curve corresponds to the gap minima $|\Delta|_{\min}/T_{c}$. From Eq. (\ref{eq_op}) it is easy to see that $|\Delta|_{\max}=\Delta_{0}$ and $|\Delta|_{\min} = \max\{ \Delta_{0}(1-2x),0\}$.
The maximum of the gap function $|\Delta|_{\max}$, along with values of the gap at saddle points, is related to a cusp or a logarithmic singularity in the density of states. When the $s$-wave component is dominate over the $g$-wave, {\it i.e.} $x<0.5$, the minimum value of the gap function is $|\Delta|_{\min}=\Delta_{0}(1-2x)$. Notice that $|\Delta|_{\min}$ is not a simple linear function of $x$, since $\Delta_{0}$ also varies with $x$ in a complicated way. $|\Delta|_{\min}$ appears as a gap in the density of states. When the $g$-wave component starts to dominate, {\it i.e.} $x \ge 0.5$, $|\Delta|_{\min}$ is zero, and the density of states does not show any gap at all. These features are shown in Fig. 3.

The quasiparticle's density of states in the superconducting state can be calculated from the following equation.
\begin{equation}
N_{s}(E) = N(0){\rm Re} \bigg< \frac{|E|}{\sqrt{E^{2} + |\Delta(\hat{k})|^{2}}} \bigg>_{FS}.
\label{eq_dos1}
\end{equation}
When there is only an $s$-wave component, the gap function has no structure in $\hat{k}$-space and the density of states becomes well-known BCS density of states curve which is the solid curve in Fig. 3. As the $g$-wave component grows, the gap function develops anisotropy in $\hat{k}$-space and the density of states has more structures. For $x<0.5$, there does not exist any state at energies below $|\Delta|_{\rm min}$, {\it i.e.}, $E<\Delta_{0}(1-2x)$, and it is shown for $x=0.3$ case in Fig. 3. For $x \ge 0.5$ the density of states shows a gapless feature. In the low energy region, we observe that the density of states has power-law dependence on energy $E$, mostly linear dependence. The curve for $x=0.8$ shows linear dependence on $E$. It is easily understood from the fact that there exist line nodes for $x>0.5$. Interesting observation is that the curve for $x=0.5$ also shows linear dependence on $E$ in the low energy regime. Notice that there are four point nodes along the equator in the gap function, when $x=0.5$ as shown in Fig. 1(c). We would usually  expect $N_{s}(E) \propto E^{2}$ dependence in the presence of point nodes.\cite{gor,sig,paul,hskim02} To understand the linear dependence coming from point nodes, we need to have a closer look at the structure of the point node. Expending the gap function near the nodes, we have
\begin{equation}
\Delta(\hat{k}) \sim \Delta_{0} [(\theta - \theta_{0})^{2} + 4(\phi - \phi_{0})^{2}],
\label{eq_gapex}
\end{equation}
where $(\theta_{0}, \phi_{0})$ is the position of the nodes, {\it i.e.}, $\theta_{0} = \frac{\pi}{2}$, $\phi_{0}=\pm \frac{\pi}{4}$, $\pm\frac{3\pi}{4}$.
Equation (\ref{eq_dos1}) tells us that, for small $E$, only contribution on $N_{s}(E)$ comes from near the nodes.
\begin{equation}
\small{
N_{s}(E) \sim N(0)\int_{E>|\Delta(\hat{k})|} \frac{d \hat{k}}{4\pi} \frac{|E|}{\sqrt{E^{2}-\Delta_{0}^{2}[(\theta - \theta_{0})^{2} + 4(\phi - \phi_{0})^{2}]^{2}}}
}
\label{eq_dos2}
\end{equation}
Although the integral looks complicated, by rescaling and by changing the variables few times, it can be transformed into a simpler form.
\begin{equation}
N_{s}(E) \sim z \frac{N(0)}{4}\int_{0}^{E>\Delta_{0}\eta^{2}} d\eta  \frac{\eta|E|}{\sqrt{E^{2}-\Delta_{0}^{2}\eta^{4}}},
\label{eq_dos3}
\end{equation}
where $z$ is the number of the point nodes, which is 4 in our case. In turn,
\begin{equation}
N_{s}(E) \sim N(0) \frac{|E|}{2\Delta_{0}}\int_{0}^{1} dy \frac{1}{\sqrt{1-y^{2}}} = N(0) \frac{\pi}{4} \frac{|E|}{\Delta_{0}}.
\label{eq_dos4}
\end{equation}
%
%
\begin{figure}
\includegraphics[width=8.0cm]{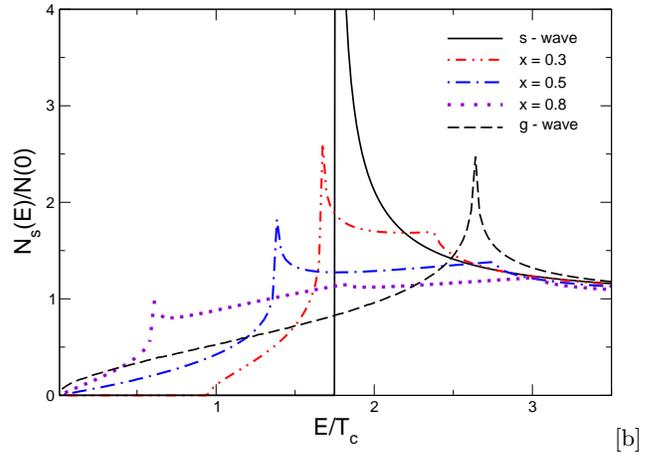}[b] 
\caption{\label{fig:dos} Density of states for various weighting parameters $x$.}
\end{figure}
%
Thus, the point nodes, present in $x=0.5$ case, produce linear $E$ dependence rather than $E^{2}$ dependence in $N_{s}(E)$. In fact, form of Eq. (\ref{eq_dos3}) tells us that it is equivalent to the case of an order parameter with quadratically vanishing polar point nodes, which is of the form $\Delta(\theta) \sim \Delta_{0} \theta^{2}$ near the nodes. From our previous work,\cite{hskim02} we know that $N_{s,low}(E) \propto E^{2/n}$ when the order parameter has
point polar node vanishing with n-th power, {\it i.e.} $\Delta(\theta) \sim \Delta_{0} \theta^{n}$ near the node.
Therefore, we can understand the linear dependence from the point nodes.
When $x=1$, the gap function has only $g$-wave component which has eight line nodes along the azimuthal angle and two point nodes at poles. Contributions on the density of states $N_{s}(E)$ from the line nodes would be linear, $N_{s}(E) \propto |E|$. The contribution from the point node would dominate over those from line nodes, since the gap function vanishes as $\propto \theta^{4}$ at the poles and it would give $N_{s}(E) \propto |E|^{1/2}$ dependence.
However, the nodal structure in the $g$-wave is little more complicated, since the line nodes cross the point nodes.
The nodal structure is too complicated to extract the energy dependence of $N_{s}(E)$ from, by using the leading order calculation. The numerical calculation gives us $N_{s}(E) \propto E^{0.53}$ in the low energy regime. It is also shown in Fig. 3.

Electronic contributions on the specific heat can be calculated from the following equation.\cite{tinkham}
\begin{equation}
C_{es} =
-2 \beta^{2} \Sigma_{k} E_{k}\left(E_{k}+\beta\frac{\partial E_{k}}
{\partial \beta} \right) \frac{\partial f_{k}}{\partial (\beta E_{k})},
\end{equation}
where $\beta=1/T$, and $f_{k}$ is the Fermi function,
\begin{equation}
f_{k} = \frac{1}{e^{\beta E_{k}}+1},
\end{equation}
and $E_{k}$ is the quasiparticle's exitation energy,
\begin{equation}
E_{k}= \sqrt{\epsilon^{2}(k)+|\Delta(k)|^{2}}.
\end{equation}
%
%
\begin{figure}
\includegraphics[width=8.0cm]{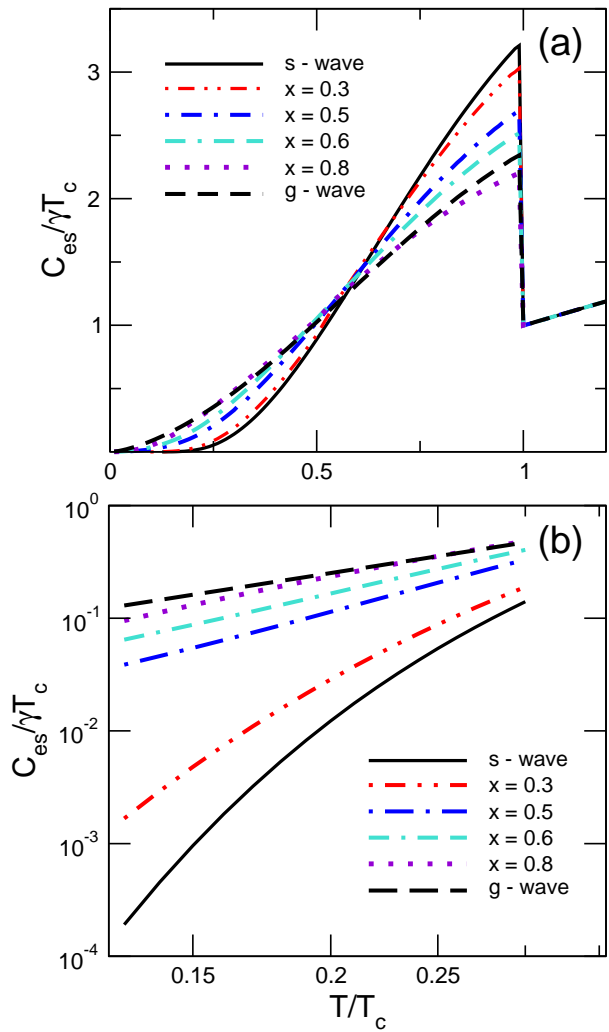}
\caption{\label{fig:sp} (a) Specific heat as a function of temperature for various weighting parameters $x$ It is normalized by the normal state value at $T_{c}$. Note that the specific heat in the normal state $C_{en,low} \sim \gamma T$. (b) the same as in (a), but in log-log scales.}
\end{figure}
%
The specific heat is plotted for various weighting parameters $x$ in Fig. 4(a) and 4(b). The solid curves in Fig. 4 correspond to the BCS case, namely, the $s$-wave case. In the low temperature region, the specific heat of the BCS case is known to follow the exponential temperature dependence, more precisely, $C_{es} \propto T^{-3/2}\, {e}^{-\Delta/T}$.
The dash-dot-dot curves are for $x=0.3$. A small $g$-wave component is mixed in and, yet, there exists a gap of size $\sim 0.95 T_{c}$ in the density of states. This finite size gap induces the exponential temperature dependence similar to the BCS case. Due to the similar exponential dependence, curves for the two cases almost fall on each other in Fig. 4(a). The difference is exposed in Fig. 4(b), a log-log plot. Figure 4(b) shows that the $s$-wave and $x=0.3$ curves are distinctly different from the other curves which are, in fact, almost straight lines. The other curves correspond to the case with nodes in the gap. When there are nodes in the gap function on the Fermi surface, it results in the power-law temperature dependence in the specific heat in the low temperature regime. It is directly related to the power-law dependence of the quasiparticle's density of states. At low temperatures the gap amplitude is almost constant in temperatures, and the specific heat can be simplified as
\begin{equation}
C_{es} \sim 2  \beta^{2} \int_{0}^{\infty} dE \, N_{low}(E) \, E^{2}\, e^{-\beta E}.
\end{equation}
Here $ N_{low}(E)$ is the density of states at low energies and it
follows some power-law in energy when there exist nodes, as discussed above.
Let us write the density of states as $N_{low}(E) \sim A E^{a}$, where $A$ and $a$ are some constants related to the detailed structure of the nodes. Then, the specific heat becomes
\begin{eqnarray}
C_{es} &\sim& 2 \beta^{2} \int_{0}^{\infty} dE \, A E^{a} \, E^{2} e^{-\beta E}, \\
&=& 2 A \Gamma(a+3) T^{a+1}.
\end{eqnarray}
Therefore, the density of state of form $N_{low}(E) \propto E^{a}$ leads us to the specific heat of form $C_{es, low}(T) \propto T^{a+1}$ in the superconducting state. This relation can be found in other literatures as well.\cite{gor,sig,paul,hskim02} This relation help us to understand the straight lines, meaning the power-law dependence, for $x=0.5, 0.6, 0.8$, and $g$-wave in Fig. 4(b). As the temperature goes down, the graphes for $x=0.5, 0.6, 0.8$ have slopes $\sim 2$, which corresponds to $C_{es, low}(T) \propto T^{2}$. The slope of the line for $g$-wave is slightly different from other lines, and it is estimated as $\sim 1.53$, meaning $C_{es, low}(T) \propto T^{1.53}$. These are consistent with the result in Fig. 3.

We have investigated evolution of various properties of an extended $s$-wave superconductor, when the order parameter varies from the $s$-wave to the $g$-wave continuously, mainly focusing on the gap amplitude, the density of states, and the specific heat. When the gap function is normalized on the Fermi surface, the gap amplitude decreases, as the anisotropy increases. When the $s$-wave dominates, {\it i.e.} the weighting parameter $x<0.5$, characteristic behaviors of the density of states and the specific heat more or less follow those of the $s$-wave. When $x=0.5$, there appear point nodes of the order parameter on the Fermi surface, and there are sudden changes in the behaviors of the density of states and the specific heat, especially in the low temperature and low energy regimes. As the weighting parameter $x$ further increases, $x>0.5$, the point nodes change into line nodes, and yet there is not any characteristic change in the superconducting properties from the case of $x=0.5$. It is shown that the point nodes of $x=0.5$ case are virtually the same as the quadratically vanishing polar point node which would produce $N_{s,low}(E) \propto |E|$. Therefore, the effect of the point nodes of $x=0.5$ case is the same as the effect of the line nodes of  $x>0.5$ case. When $x=1$, we have the $g$-wave and there are again characteristic changes in low temperatures and in low energies. These characteristic changes thoroughly come from the changes of the nodal structure. We hope this work would encourage more attentions to be paid on the superconducting properties coming from the nontrivial nodal structures.

The authors are grateful to C. H. Choi for useful discussions.
This research was supported by Basic Science Research Program through the National Research Foundation of Korea(NRF) funded by the Ministry of Education, Science and Technology (grant 2012R1A1A2006303 and 2010-0021328).

\bibliography{Kim-references201402}
%

\end{document}